\documentclass[preprint,aps,epsf,nofootinbib,tightenlines,floats,floatfix]{revtex4}
\usepackage{epsf}
\newcommand{\beq}{\begin{equation}}
\newcommand{\eeq}{\end{equation}}
\newcommand{\beqa}{\begin{eqnarray}}
\newcommand{\eeqa}{\end{eqnarray}}

\newcommand{\no}{\nonumber}
\def\lsim{\mathrel{\rlap{\lower4pt\hbox{\hskip1pt$\sim$}}
    \raise1pt\hbox{$<$}}}         
\def\gsim{\mathrel{\rlap{\lower4pt\hbox{\hskip1pt$\sim$}}
    \raise1pt\hbox{$>$}}}         

\begin{document}
\preprint{\vbox{\hbox{WIS/13/02-Jun-DPP} \hbox{hep-ph/0206064}}}

\title{Quark-Squark Alignment Revisited}
\vskip 1cm
\author{Yosef Nir and Guy Raz}
\vskip 1cm
\address{Department of Particle Physics\\ Weizmann Institute of Science\\
 Rehovot 76100, Israel} 

\vskip 3cm

\begin{abstract}
We re-examine the possibility that the solution to the supersymmetric
flavor problem is related to small mixing angles in gaugino couplings
induced by approximate horizontal Abelian 
symmetries. We prove that, for a large class of models, there is a single 
viable structure for the down quark mass matrix with four holomorphic zeros. 
Consequently, we are able to obtain both lower and upper bounds on the 
supersymmetric mixing angles and predict the contributions to various flavor 
changing neutral current processes. We find that the most likely signals for 
alignment are $\Delta m_D$ close to the present bound, significant CP violation
in $D^0-\overline{D^0}$ mixing, and shifts of order a few percent in various
CP asymmetries in $B^0$ and $B_s$ decays. In contrast, the modifications
to radiative $B$ decays, to $\varepsilon^\prime/\varepsilon$ and to
$K\to\pi\nu\bar\nu$ decays are small. We further investigate a new class of 
alignment models, where supersymmetric contributions to flavor changing 
processes are suppressed by both alignment and RGE-induced degeneracy.
\end{abstract}
\maketitle

\section{Introduction}
Quark-squark alignment (QSA) is a mechanism that suppresses supersymmetric 
contributions to flavor changing neutral current processes via small mixing 
angles in flavor changing gaugino couplings \cite{Nir:1993mx,Leurer:1993gy}. 
The alignment could be precise enough that the models are viable without 
requiring any squark degeneracy. Alignment occurs naturally in all models with 
Abelian horizontal symmetries that induce the observed hierarchy in the Yukawa 
couplings. However, to achieve small enough mixing angles in the gaugino 
couplings that are relevant to $\Delta m_K$ and $\varepsilon_K$, one has to 
carefully choose the symmetry and the charge assignments.

Existing models of alignment use holomorphic zeros in the down quark mass
matrix to achieve small enough mixing angles between the first two generations.
In section II we re-examine the allowed structures for this mass matrix. We 
prove that there is a single structure (that is, a unique set of holomorphic 
zeros) that gives phenomenologically viable mixing angles. The unique structure
of $M^d$ gives this framework a strong predictive power: We are able to derive 
both lower and upper bounds on the parametric suppression of the supersymmetric
mixing angles.

The most interesting prediction of models of quark-squark alignment is that
the mass difference in the neutral $D$ system, $\Delta m_D$, should be close
to the experimental bound. (A more refined version of this statement is given
in section III.) Furthermore, $D^0-\overline{D^0}$ mixing could be
CP violating. While recent analyses suggest that the Standard Model 
contribution to $\Delta m_D$ could also be large \cite{Falk:2001hx}, CP 
violation in the mixing will provide unambiguous evidence for new physics.
Recently, there has been much progress in the search for $D^0-\overline{D^0}$ 
mixing. No signal has been found, and the bounds on the mixing parameters
have improved.  In section III we examine the implications of these improved
bounds on the viability of QSA models. It is important here that the 
experimental results on $D^0-\overline{D^0}$ mixing are analyzed allowing
for CP violation. We thus use the results of ref. \cite{Raz:2002ms} where
the impact of weak (and strong) phases on the interpretation of the
experimental bounds was taken into account.

The framework of alignment has a strong predictive power also for the mixing
angles related to $B^0-\overline{B^0}$ mixing, $B_s-\overline{B_s}$ mixing
and $b\to X\gamma$ decays. We analyze these predictions in section IV. The 
implications for $K$ physics $-$ $\varepsilon^\prime/\varepsilon$ and $K\to
\pi\nu\bar\nu$ decays $-$ are discussed in section V.

Another basic assumption made in the literature is that the {\it only}
restriction on the soft supersymmetry breaking terms comes from the selection
rules related to the small breaking of the horizontal symmetry. In particular,
it was assumed that there is no degeneracy among squark masses, that is,
$\Delta m^2/m^2={\cal O}(1)$. This assumption may, however, be questioned.
It is perhaps more plausible that this situation holds at a high energy
scale, where the soft supersymmetry breaking terms are induced. But then,
renormalization group evolution (RGE) would give a universal contribution
to squark masses and lead to some degree of degeneracy. In section VI we
examine various aspects of `high energy alignment': we estimate the size
of the effect and its consequences for the constraints on mixing angles
and for model building.  

Future prospects for finding evidence for the alignment mechanism or
for excluding it are discussed in section VII.

\section{Supersymmetric Mixing Angles}
The size of supersymmetric flavor violation depends on the overall scale
of the soft supersymmetry breaking terms, on mass degeneracies between
sfermion generations, and on the mixing angles in gaugino couplings. Within
the framework of alignment, mixing angles play a significant role. For most
of our purposes here, we can make the approximation that the mixing between 
$\tilde q_L$, the superpartners of the left-handed quarks, and $\tilde q_R$,
the superpartners of the right-handed quarks, is small. Then there are four
relevant $3\times3$ mixing matrices in the quark-squark sector, which we denote
by $K^d_L$, $K^d_R$, $K^u_L$ and $K^u_R$. 

Consider, for example, the matrix elements $(K^d_L)_{ij}$ which parametrize the 
$\tilde g-(d_L)_i-(\tilde d_L)_j$ couplings. Given the down quark mass matrix
in the interaction basis, $M^d$, we define the diagonalizing matrices, $V^d_L$
and $V^d_R$, according to
\beq\label{diaqua}
V^d_L M^d V^{d\dagger}_R={\rm diag}(m_d,m_s,m_b).
\eeq
Given the the mass-squared matrix for the $\tilde d_L$ squarks,
$\tilde M^{2d}_{LL}$, we can obtain the diagonalizing matrix $\tilde V^d_L$:
\beq\label{diasqu}
\tilde V^d_L \tilde M^{2d}_{LL} \tilde V^{d\dagger}_L={\rm diag}
(m^2_{\tilde d_1},m^2_{\tilde d_2},m^2_{\tilde d_3}).
\eeq
Then we have
\beq\label{keqvv}
K^d_L=V_L^d\tilde V_L^{d\dagger}.
\eeq
In this chapter we derive predictions for the flavor changing elements
of the $K^q_M$ matrices in the framework of alignment.

\subsection{The Down Quark Mass Matrix}
If the only suppression of supersymmetric flavor violation is related to 
alignment, then the constraints from $K^0-\overline{K^0}$ mixing ($\Delta m_K$ 
and $\varepsilon_K$) require that the relevant supersymmetric mixing angles are
much smaller than the corresponding CKM angle:
\beq\label{alignds}
|(K^d_{L})_{12}|,\ |(K^d_{R})_{12}|\ll|V_{us}|=\lambda.
\eeq
In models where alignment is induced by an Abelian horizontal symmetry, such
a situation can be achieved by having holomorphic zeros in the down quark
mass matrix.

We would like to argue that, for a large class of alignment models based on 
Abelian horizontal symmetries, there is a unique structure for the down quark 
mass matrix which is consistent with (\ref{alignds}) and with the known values 
of the quark flavor parameters (masses and mixing angles):\footnote{For related
studies, see \cite{Berger:2000zj,Barr:1997ph}.}
\beq\label{uniquemd}
M^d\sim\pmatrix{m_d&0&m_bV_{ub}\cr 0&m_s&m_bV_{cb}\cr 0&0&m_b\cr}.
\eeq
(The `$\sim$' sign here and below means that there is an arbitrary coefficient
of order one, which we do not write explicitly, in each entry.) 
We will now prove this statement and spell out our assumptions along the way.

In order that (\ref{alignds}) is satisfied, we must have 
$|(V^d_L)_{12}|,\ |(V^d_R)_{12}|\ll\lambda$. These
matrix elements can be expressed in terms of the entries of $M^d$
\cite{Hall:1993ni,Leurer:1993gy}. We define:
\beqa\label{defyd}
y^d_{i1}&=&{M^d_{i1}\over\sqrt{|M^d_{22}|^2+|M^d_{33}|^2}},\no\\
y^d_{i2}&=&{M^d_{i2}M^d_{33}-M^d_{i3}M^d_{32}\over|M^d_{22}|^2+|M^d_{33}|^2},
\no\\
y^d_{i3}&=&{M^d_{i3}M^{d*}_{33}+M^d_{i2}M^{d*}_{32}\over
|M^d_{22}|^2+|M^d_{33}|^2}.
\eeqa
Then the relevant contributions to the matrix elements are given as follows:
\beqa\label{sonetwod}
(V^d_L)_{12}&=&{y^d_{12}\over y^d_{22}}+{y^d_{11}y^{d*}_{21}\over|y^d_{22}|^2}
,\no\\
(V^d_R)_{12}&=&{y^{d*}_{21}\over y^{d*}_{22}}+{y^{d*}_{11}y^d_{12}\over
y^{d*}_{22}}-{y^{d*}_{31}y^{d*}_{23}\over y^{d*}_{22}}.
\eeqa
To sufficiently suppress these mixing angles while providing acceptable
values for the down quark masses, the following conditions are necessary
\cite{Eyal:1999gk}:
\begin{itemize}
\item $M^d_{12}=0$;
\item $M^d_{21}=0$;
\item $M^d_{31}=0$ or $M^d_{23}=M^d_{32}=0$;
\item $M^d_{32}=0$ or $M^d_{13}=0$.
\end{itemize}
But not all the ways to satisfy these conditions can be realized in models
of Abelian horizontal symmetries. In particular, we will now prove that in
a large class of models we can have neither $M^d_{13}=0$ nor $M^d_{23}=0$.

We consider models with Abelian symmetries of the type $U(1)_1\times U(1)_2
\times\cdots \times U(1)_n$. Each $U(1)_i$ subgroup is broken by a small 
parameter $\epsilon_i$. It is convenient to express all $\epsilon_i$'s as 
powers of $\lambda$, $\epsilon_i\sim\lambda^{n_i}$ ($n_i>0$). We emphasize 
that there is no loss of generality in doing so.
Each matter supermultiplet $\Phi$ carries horizontal charges $H_i(\Phi)$, 
$i=1,\ldots,n$. Here $\Phi$ stands for any of the quark doublet superfields
$Q_i$, the singlet anti-up superfields $\bar u_i$, the singlet anti-down 
superfields $\bar d_i$ and the Higgs superfields $\phi_u$ and $\phi_d$. We use 
the freedom that comes from the $U(1)_Y\times U(1)_B \times U(1)_{\rm PQ}$ 
symmetry of the Yukawa sector to set $H_i(Q_3)=H_i(\phi_u)=H_i(\phi_d)=0$ 
without loss of generality. It is also convenient to define an effective
charge of a field, $H(\Phi)=\sum_i n_iH_i(\Phi)$. 
Then the selection rules for the entries in $M^q$ ($q=d,u$) are as follows:

(i) If, for all $i$, $H_i(Q_j)+H_i(\bar q_k)\geq0$ then $M^q_{jk}=\langle
\phi_q\rangle\lambda^{H(Q_j)+H(\bar q_k)}$.

(ii) If, for some $i$,  $H_i(Q_j)+H_i(\bar q_k)<0$ then $M^q_{jk}=0$.

We assume that $m_t/\langle\phi_u\rangle={\cal O}(1)$, namely it is not
parametrically suppressed.\footnote{The alignment model of ref. 
\cite{Chua:2001dd} takes $m_t$ to be parametrically suppressed and therefore is
not subject to our analysis. Similarly, neither the mass matrix structures nor 
the phenomenological consequences proposed in refs. 
\cite{Arhrib:2001jg,Diaz-Cruz:2001gf} are possible in our framework.}
Then we must have $H_i(\bar u_3)+H_i(Q_3)=0$
for all $i$. We also must have $M^d_{33}\simeq m_b$ which means that
$H_i(\bar d_3)+H_i(Q_3)\geq 0$ for all $i$. These two conditions together
imply that $H_i(\bar d_3)\geq H_i(\bar u_3)$. Then it is simple to see that
if $M^d_{i3}=0$, we necessarily have also $M^u_{i3}=0$. But if $M^d_{23}=
M^u_{23}=0$ we would obtain $|V_{cb}|\ll\lambda^2$. We conclude that 
we must not have $M^d_{23}=0$ and that, therefore, we must have $M^d_{31}=0$.
But if $M^d_{31}=M^d_{13}=M^u_{13}=0$ we would obtain $|V_{td}|\ll\lambda^3$.
We conclude that we must not have $M^d_{13}=0$ and that, therefore, we must
have $M^d_{32}=0$.  This completes the proof to our statement
that the only viable down quark mass matrix within our framework and
assumptions is that of eq. (\ref{uniquemd}).

\subsection{The Supersymmetric Mixing Angles}
In the framework of alignment one assumes that there are no fine-tuned 
relations between ${\cal O}(1)$ coefficients. This means that we can use eq.
(\ref{keqvv}) to estimate $(K^d_{L})_{ij}$:
\beq\label{lowkll}
(K^d_{L})_{ij}\sim{\rm max}\left[(V_L^d)_{ij},(\tilde V_L^d)_{ji}\right].
\eeq

The uniqueness of the mass matrix $M^d$ of eq. (\ref{uniquemd}) implies that
the parameteric suppression of all entries of the diagonalizing matrices
$V^d_{L,R}$ is known within our framework:
\beq\label{dimaali}
V_L^d\sim\pmatrix{1&\lambda^5&\lambda^3\cr \lambda^5&1&\lambda^2\cr
\lambda^3&\lambda^2&1\cr},\ \ \  
V_R^d\sim\pmatrix{1&\lambda^7&\lambda^7\cr \lambda^7&1&\lambda^4\cr
\lambda^7&\lambda^4&1\cr}.
\eeq 
From eq. (\ref{lowkll}) we conclude that the values of the various entries in 
$V_L^d$ and $V_R^d$ given in eq. (\ref{dimaali}) constitute lower bounds on the
corresponding entries in, respectively, $K^d_{L}$ and $K^d_{R}$. In other 
words, the parametric suppression of $(K^d_{M})_{ij}$ is at most as strong as 
that of $(V^d_M)_{ij}$ in eq. (\ref{dimaali}).

We would now like to estimate the diagonalizing matrices for the squark
mass-squared matrices. The selection rules for the diagonal block are simple:

(i) For the LL block, $(\tilde M^2_{LL})_{jk}\sim\tilde m^2_Q\lambda^{
\sum_{i=1}^n n_i|H_i(Q_j)-H_i(Q_k)|}$ (for both down and up squarks).

(ii) For the RR block of the down sector, $(\tilde M^{2d}_{RR})_{jk}\sim
\tilde m^2_D\lambda^{\sum_{i=1}^n n_i|H_i(\bar d_j)-H_i(\bar d_k)|}$.

(iii) For the RR block of the up sector, $(\tilde M^{2u}_{RR})_{jk}\sim
\tilde m^2_U\lambda^{\sum_{i=1}^n n_i|H_i(\bar u_j)-H_i(\bar u_k)|}$.

(We here allow for the possibility that the typical mass-squared scale is
different for each of the three sectors. In most cases we will assume that
there is a single mass scale that characterizes all soft supersymmetry
breaking terms and denote this scale by $\tilde m$.)
The interesting point here is that one can find upper bounds on the 
off-diagonal elements of the diagonalizing matrices in terms of the
quark flavor parameters, that is, the CKM angles and the quark masses.
The latter can be written in terms of the effective charges:
\beqa\label{physobs}
|V_{ij}|&\sim&\lambda^{|H(Q_i)-H(Q_j)|},\no\\
m_{d_i}/m_{d_j}&\sim&\lambda^{H(Q_i)+H(\bar d_i)-H(Q_j)-H(\bar d_j)},\no\\
m_{u_i}/m_{u_j}&\sim&\lambda^{H(Q_i)+H(\bar u_i)-H(Q_j)-H(\bar u_j)}.
\eeqa
Then we get the following bounds (here $i\leq j$):
\beqa\label{upbotiv}
|(\tilde V^q_L)_{ij}|&\lsim&|V_{ij}|,\no\\
|(\tilde V^q_R)_{ij}|&\lsim&{1\over|V_{ij}|}\ {m_{q_i}\over m_{q_j}}.
\eeqa

There are cases in which one can derive an upper bound on 
$|(\tilde V^q_M)_{ij}|$ that is stronger than those in eq. (\ref{upbotiv}).
These are the cases when a related entry in the down quark mass matrix
is a holomorphic zero. For example, since $M^d_{31}=0$, the upper bound
on $|(\tilde V^d_R)_{13}|$ in eq. (\ref{upbotiv}), $|(\tilde V^d_R)_{13}|\lsim
(m_d/m_b)/|V_{ub}|\sim\lambda$, is never saturated and a stronger bound holds.
We now derive this bound. Our starting point is the application of the 
selection rule to this specific case,
\beq\label{drot}
|(\tilde V^d_R)_{13}|\sim \lambda^{\sum_{i=1}^n 
n_i|H_i(\bar d_1)-H_i(\bar d_3)|}.
\eeq
The source of the upper bound in eq. (\ref{upbotiv}) is the inequality
\beq\label{upboot}
\sum_{i=1}^n n_i|H_i(\bar d_1)-H_i(\bar d_3)|\geq
\sum_{i=1}^n n_i[H_i(\bar d_1)-H_i(\bar d_3)].
\eeq
For the upper bound in eq. (\ref{upbotiv}) to be saturated, eq. (\ref{upboot}) 
should become an equality.
That would imply that $H_i(\bar d_1)-H_i(\bar d_3)\geq0$ for all $i$. As we
mentioned before, we must have $M^d_{33}\simeq m_b$ which means that
$H_i(\bar d_3)+H_i(Q_3)\geq 0$ for all $i$. The combination of the two
requirements gives $H_i(\bar d_1)+H_i(Q_3)\geq 0$ for all $i$. But then
$M^d_{31}\neq0$, in contradiction to eq. (\ref{uniquemd}). The minimal
extra suppression of $|(\tilde V^d_R)_{13}|$ compared to the upper bound in
eq. (\ref{upbotiv}) is by two powers of the largest among the small parameters
$\epsilon_i$, that is,
\beq\label{upbot}
|(\tilde V^d_R)_{13}|\lsim{m_d\over|V_{ub}|m_b}\epsilon_{\rm max}^2,\ \ \ 
\epsilon_{\rm max}\equiv\max_i(\epsilon_i).
\eeq
In particular, if $\epsilon_i\lsim\lambda$ for all $i$, then
$|(\tilde V^d_R)_{13}|\lsim\lambda^3$. Together with eq. (\ref{dimaali}),
we obtain:
\beq\label{krot}
\lambda^7\lsim|(K^d_{R})_{13}|\lsim\lambda^3.
\eeq

Similar considerations apply to other supersymmetric mixing angles.
Within the up sector, the structure of the mass matrix is less restricted.
The only strict requirements are that the eigenvalues of $M^u$ would be
$(m_u,m_c,m_t)$ and that, given that the Cabibbo mixing is not induced by
the diagonalization of $M^d$, we should have $|(V_L^u)_{12}|=|V_{us}|$.
(In addition, we must have $|(V_L^u)_{13}|\lsim|V_{ub}|$ and
$|(V_L^u)_{23}|\lsim|V_{cb}|$.)
These requirements are enough to find constraints on the $|(K^u_{M})_{12}|$
mixing angles. The bounds on various mixing angles in our framework of
alignment are given in Table I.

\begin{table}[tbh]
\begin{tabular}{c|c|c}
Mixing Angle  &  Lower Bound  & Upper Bound    \\ \hline\hline
$(K^d_{L})_{12}$ & $V_{ub}V_{cb}(\sim\lambda^5)$ & 
$V_{us}\epsilon_{\rm max}^2(\sim\lambda^3)^\ddag$ \\
$(K^d_{R})_{12}$ & ${m_d\over m_s}V_{ub}V_{cb}(\sim\lambda^7)$ & 
${m_d\over m_sV_{us}}\epsilon_{\rm max}^2(\sim\lambda^3)^\ddag$ \\ \hline
$(K^d_{L})_{13}$ & $V_{ub}(\sim\lambda^3)$ & $V_{ub}(\sim\lambda^3)$ \\
$(K^d_{R})_{13}$ & ${m_d\over m_b}V_{ub}(\sim\lambda^7)$ & 
${m_d\over m_b V_{ub}}\epsilon_{\rm max}^2(\sim\lambda^3)$ \\ \hline
$(K^d_{L})_{23}$ & $V_{cb}(\sim\lambda^2)$ & $V_{cb}(\sim\lambda^2)$ \\
$(K^d_{R})_{23}$ & ${m_s\over m_b}V_{cb}(\sim\lambda^4)$ & 
${m_s\over m_bV_{cb}}\epsilon_{\rm max}^2(\sim\lambda^2)$ \\ \hline
$(K^u_{L})_{12}$ & $V_{us}\sim\lambda$ & $V_{us}(\sim\lambda)$ \\
$(K^u_{R})_{12}$ & ${m_u\over m_c}V_{us}(\sim\lambda^4)$ & 
${m_u\over m_c|V_{us}|}(\sim\lambda^2)^\ddag$ \\ 
\end{tabular}\vspace*{4pt}
\caption{Bounds on supersymmetric mixing angles in models of alignment.
The estimates in powers of $\lambda\sim0.2$ refer to our evaluation of the
quark mass ratios in powers of $\lambda$ and to $\epsilon_{\rm max}\equiv
\max_i(\epsilon_i)\lsim\lambda$.
$^\ddag$\,In viable models these mixing angles are set to be smaller than the
formal upper bounds so that the phenomenological bounds on the products 
$(K_L^q)_{12}(K_R^q)_{12}$ ($q=u,d$) are satisfied.}
\end{table}

\section{$D$ Physics}
The most promising way to find evidence for quark-squark alignment is
through CP violation in $D^0-\overline{D^0}$ mixing. The best way to
exclude a large class of alignment models is by improving the bounds
on $D^0-\overline{D^0}$ mixing. The most important quantity here is the 
dispersive part of the $D^0-\overline{D^0}$ mixing amplitude, $M_{12}^D$.  
To constrain the supersymmetric flavor parameters, we need to find the 
phenomenological bounds on this transition amplitude. The analysis is not 
straightforward because the possible presence of strong phases and of weak 
phases in the relevant decay processes complicates the relation between 
$M_{12}$ and the experimentally measured parameters. A careful analysis
was performed in ref. \cite{Raz:2002ms} with the result\footnote{In the 
literature, the effects of weak and strong phases on the interpretation of 
searches for $D^0-\overline{D^0}$ mixing are often ignored. Consequently, a 
stronger bound is often quoted. See \cite{Raz:2002ms} for details.}
\beq\label{motexp}
|M_{12}^D|\leq6.2\times10^{-11}\ MeV\ (95\%\ {\rm CL}).
\eeq
In the next subsection we interpret this bound in the framework of
supersymmetric models with quark-squark alignment.

\subsection{Mixing Angle Constraints Without Squark Degeneracy}
Supersymmetric box diagrams with intermediate gauginos and squarks contribute
to neutral meson mixing. It is our purpose in this subsection to estimate the 
supersymmetric contribution to $M_{12}^D$ in the framework of quark-squark
alignment models and to compare it to the experimental bound (\ref{motexp}). 

The size of the contribution depends on the masses of the intermediate 
particles and on the mixing angles in the gaugino couplings to quarks and 
squarks. The interest in $D^0-\overline{D^0}$ mixing lies in the fact that
alignment models predict the value of one relevant mixing angle:
\beq\label{kotuo}
|(K^u_L)_{12}|\simeq\lambda.
\eeq
Here $\lambda\equiv|V_{us}|=0.22$ is the Wolfenstein parameter. The mixing
angle $(K^u_L)_{12}$ gives the coupling of the gluino (or a neutralino) to
a left-handed up quark and a `left-handed' charm squark. Then one
can calculate the contribution to $M_{12}^D$ in terms of the three relevant
masses, $m_{\tilde g}$, $\tilde m_2$ and $\tilde m_1$ (where the latter
are, respectively, the masses of $\tilde c_L$ and $\tilde u_L$). 

\begin{figure}[t]
\centerline{\epsfxsize=0.5\textwidth \epsfbox{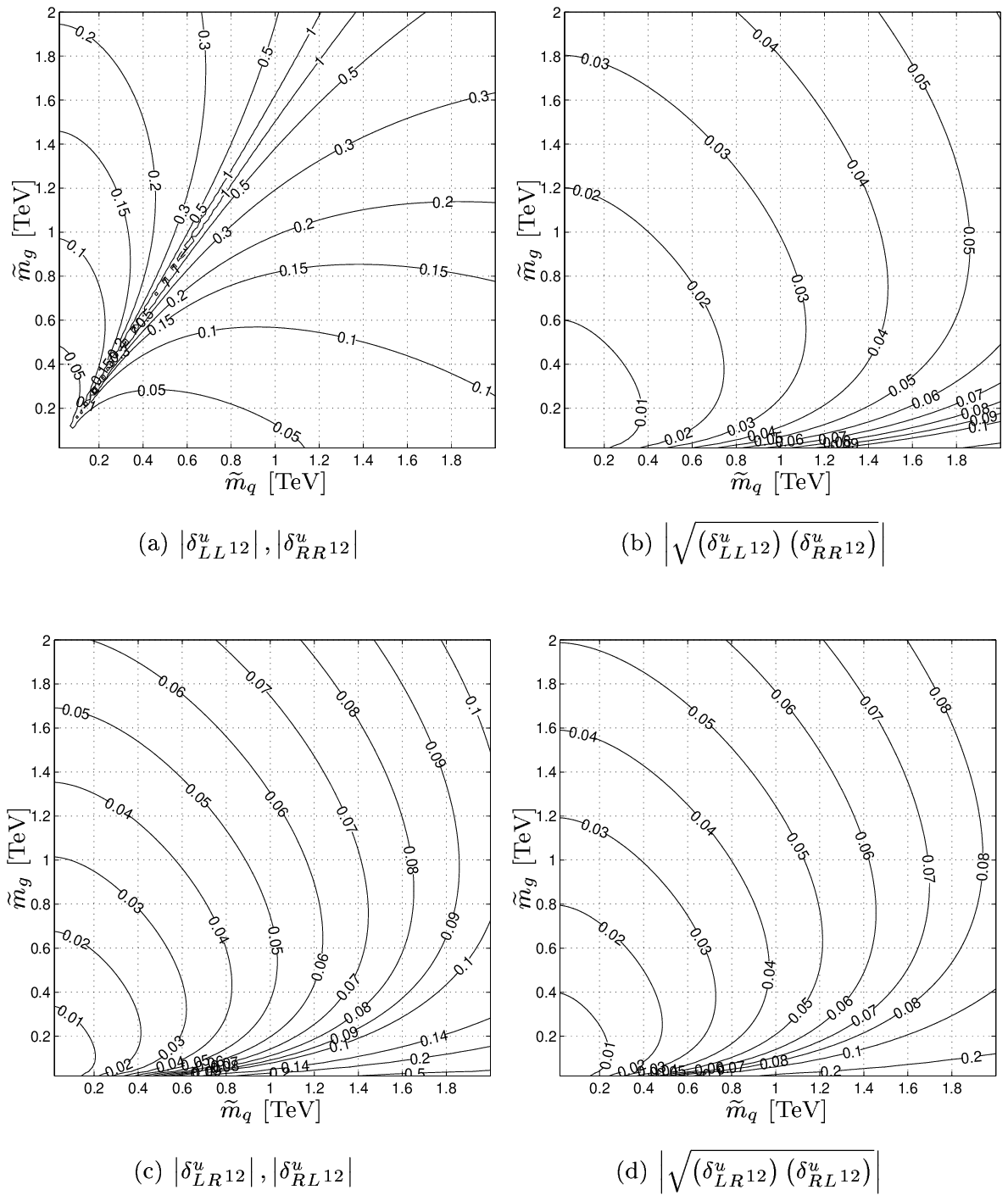}}
\caption{Constraints on flavor changing mass insertions from 
$D^0-\overline{D^0}$ mixing as a function of the gluino mass $m_{\tilde g}$
and of the average squark mass $\tilde m_Q$.}
\label{falign1}
\end{figure}

One often calculates the supersymmetric contributions to neutral meson mixing
in the mass insertion approximation (MIA). This is equivalent to Taylor
expanding around a common squark mass $\tilde m_Q$ and keeping only the
leading term in $\Delta\tilde m^2_{21}/\tilde m_Q^2$, where
\beqa\label{deftil}
\tilde m_Q&=&\frac{1}{2}(\tilde m_2+\tilde m_1),\no\\
\Delta\tilde m^2_{21}&=&(\tilde m_2^2-\tilde m_1^2).
\eeqa
(The particular choice of $\tilde m_Q$ in eq. (\ref{deftil}) is explained in 
ref. \cite{Raz:2002zx}.) It is convenient to define the following dimensionless 
quantity:
\beq\label{defdel}
(\delta^u_{LL})_{12}\equiv{(V^u_L \tilde M^{2u}_{LL} V^{u\dagger}_L)_{12}\over
\tilde m_Q^2} \sim (K^{u}_{L})_{12}{\Delta\tilde m^2_{21}\over\tilde m^2_Q}.
\eeq
In the second equation we assumed that the terms related to $(K^u_L)_{13}
(K^u_L)_{23}$ can be neglected and that, furthermore, the diagonal matrix
elements, $(K^u_L)_{ii}$, are not parametrically suppressed. These assumptions
are always valid in our framework of Abelian horizontal symmetries.
The leading contribution in the MIA depends on $m_{\tilde g}$, $\tilde m_Q$
and $(\delta^u_{LL})_{12}$. The MIA result for the contributions to $M_{12}^D$
involving $\tilde c_L$ and $\tilde u_L$ is given by
\cite{Gabbiani:1996hi}
\beq\label{motmia}
M_{12}^D={\alpha_s^2m_DB_Df_D^2\eta_D\over\tilde m_Q^2}\left[
{11\over108}\tilde f_6(m_{\tilde g}^2/\tilde m_Q^2)+
{1\over27}{m_{\tilde g}^2\over\tilde m_Q^2}f_6(m_{\tilde g}^2/\tilde m_Q^2)
\right]\left[(\delta^u_{LL})_{12}\right]^2,
\eeq
where
\beqa\label{deffsix}
f_6(x)&=&{6(1+3x)\ln x+x^3-9x^2-9x+17\over6(1-x)^5},\no\\
\tilde f_6(x)&=&{6x(1+x)\ln x-x^3-9x^2+9x+1\over3(1-x)^5}.
\eeqa
Similarly, one can find the contributions that are porportional to 
$\left[(\delta^u_{RR})_{12}\right]^2$, $(\delta^u_{LL})_{12}
(\delta^u_{RR})_{12}$, $\left[(\delta^u_{LR})_{12}\right]^2$,
$\left[(\delta^u_{RL})_{12}\right]^2$, and $(\delta^u_{LR})_{12}
(\delta^u_{RL})_{12}$. (Generalizing eq. (\ref{defdel}), one defines
$\delta^q_{MN}\equiv V_M^q\tilde M^{2q}_{MN}V_N^{q\dagger}/\tilde m^2$.)
Requiring that each of these contributions separately
is smaller than our bound (\ref{motexp}) gives an upper bound on each
of the $(\delta^u_{MN})_{12}$ combinations. These bounds are shown in Fig. 
\ref{falign1}. For example, with $m_{\tilde g}=\tilde m_Q=1\ TeV$, we find:
\beq\label{dddel}
(\delta^u_{LL})_{12}\lsim0.2.
\eeq

Note that we do not take into account possible fine-tuned cancellations between
the various contributions. Such cancellations would allow weaker bounds. While 
this option goes against the spirit of our work, where we try to explain small 
numbers by parametric suppression related to approximate symmetries and not by
fine-tuning, one has to bear in mind that it is not impossible that the
bounds are violated by a factor of a few and accidental cancellation does
take place in Nature \cite{Arhrib:2001gr}.

How should we interpret constraints that are calculated with the MIA within
the framework of alignment? The answer is not simple for the following reason.
Within models of alignment, the suppression of flavor changing 
$(\delta^q_{MN})_{ij}$ comes from the smallness of the mixing angles and not 
from squark degeneracy. Actually, in the spirit of alignment models, where all 
couplings that are not suppressed by the approximate horizontal symmetry are 
expected to be of ${\cal O}(1)$, one usually further assumes that there is no 
degeneracy among the relevant squarks, that is,
\beq\label{nodeg}
{\Delta \tilde m^2_{21}\over\tilde m^2_Q}={\cal O}(1).
\eeq
But the MIA is an expansion in $\Delta\tilde m^2/\tilde m^2$ (and not in 
$\delta$). Therefore it is not necessarily a good approximation for alignment 
models. Ref. \cite{Raz:2002zx} investigated the relation between the MIA and 
exact calculations within alignment models. The conclusion is that,
in most of the parameter space, the MIA with the choice of $\tilde m_Q$ as 
in eq. (\ref{deftil}) is a good approximation for the exact result. Thus, in 
the absence of any squark degeneracy, the constraints in Fig. \ref{falign1} 
should be interpreted as an approximate upper bound on the mixing angle 
$|(K^u_L)_{12}|$. The approximation breaks only if there is a strong hierarchy 
between the two squark masses. If, on the other extreme, there is approximate 
degeneracy between the two squark masses, then the MIA constraint is (close to)
exact but it applies to $|(K^u_L)_{12}|(\Delta\tilde m^2/\tilde m^2)$. 
 
\subsection{$M_{12}^D$ with Quark-Squark Alignment}
In all models of alignment, eq. (\ref{kotuo}) holds for the mixing angle.
In the class of models considered in this section, eq. (\ref{nodeg}) is
assumed. In this class of models, the generic prediction is then that
\beq\label{aldel}
(\delta^u_{LL})_{12}\sim0.2.
\eeq
to be compared with the experimental bound of eq. (\ref{dddel}) or, more
generally, with the constraints of Fig. \ref{falign1}(a). The regions of 
parameter space where the constraint on $(\delta^u_{LL})_{12}$ is stronger 
than 0.2 are disfavored. The regions where the constraint is weaker are viable.
We can make then the following three statements:

(i) Models of quark-squark alignment where $m_{\tilde g},\tilde m_Q\gsim1\ TeV$
are consistent with the experimental constraints from $D^0-\overline{D^0}$
mixing without any squark degeneracy. 

(ii) Conversely, models where both of $m_{\tilde g}$ and $\tilde m_Q$
are much lighter than $1\ TeV$ are disfavored, unless there is some degeneracy
between the first two generations of squarks.

(iii) There is a narrow region in the $m_{\tilde g},\tilde m_Q$ plane where
various contributions to $M_{12}^D$ cancel against each other and the
supersymmetric particles could be very light without violating the bound. While
exact cancellation is unlikely, one should bear in mind that an accidental,
approximate cancellation is possible and the $TeV$ bound on the masses is
not strict.

If supersymmetry is to solve the fine tuning problem, supersymmetric masses 
should be $\lsim TeV$. The conclusion of our discussion here is then that 
models without squark degeneracy require that $|M_{12}^D|$ is close to 
present experimental bounds. 

\section{$B$ Physics}
$B^0-\overline{B^0}$ mixing and rare $B$ decays, such as the radiative
$b\to s\gamma$, are an excellent probe of supersymmetry 
\cite{Bertolini:1990if,Bertolini:1987pk}. In this section we study
the signatures of alignment in these processes.
 
\subsection{$B^0-\overline{B^0}$ mixing}
There are two important measurements that relate to $B^0-\overline{B^0}$ 
mixing. First, the mass difference between the neutral $B$ mesons is given by
\cite{Groom:in}
\beq\label{delmb}
\Delta m_B=(3.107\pm0.112)\times 10^{-10}\ MeV.
\eeq
Second, the CP asymmetry in $B\to\psi K$ decays is given by
\cite{Aubert:2001nu,Abe:2001xe}
\beq\label{apksexp}
a_{\psi K}=0.78\pm0.08.
\eeq

\begin{figure}[t]
\centerline{\epsfxsize=0.5\textwidth \epsfbox{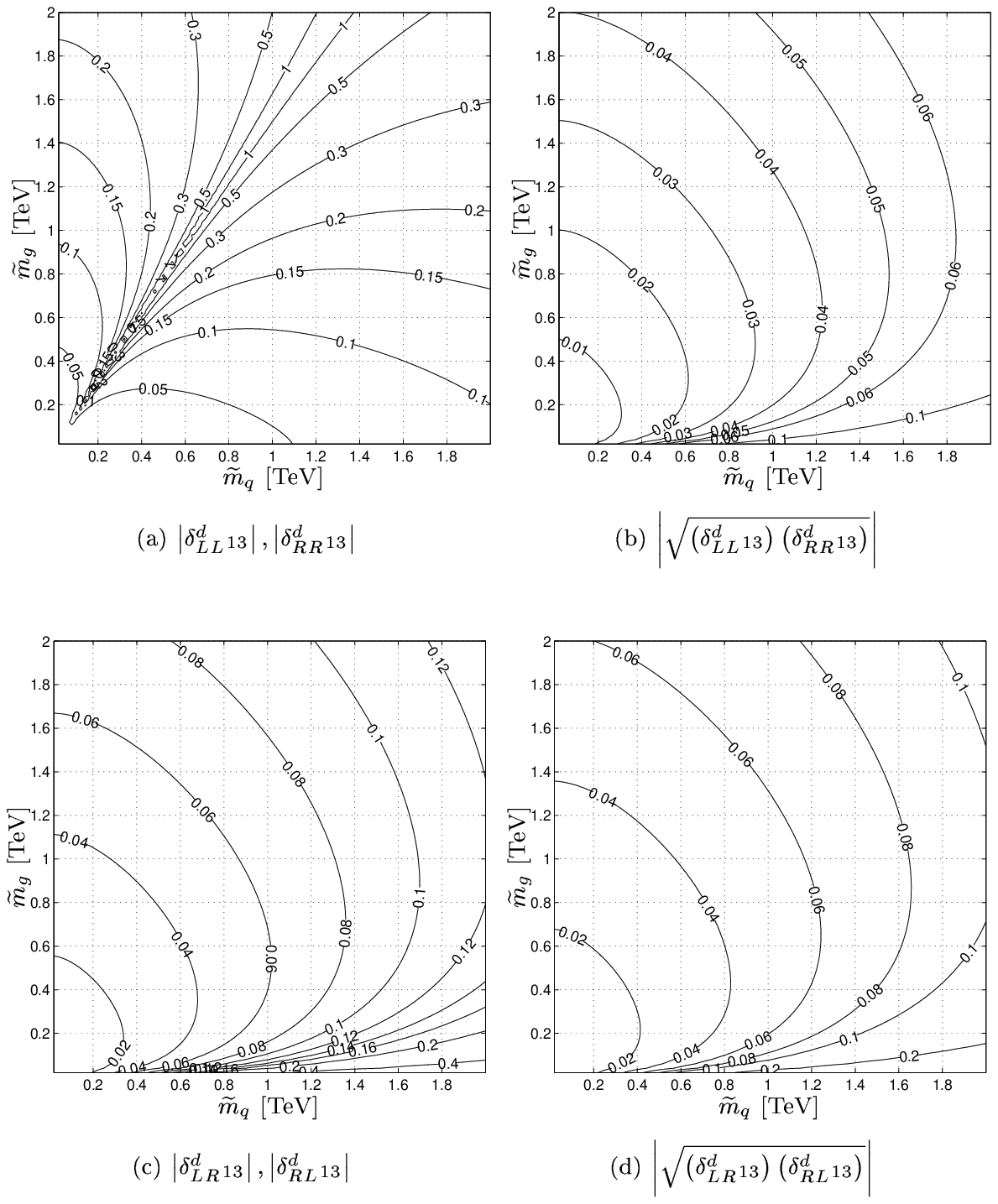}}
\caption{Constraints on flavor changing mass insertions from 
$B^0-\overline{B^0}$ mixing as a function of the gluino mass $m_{\tilde g}$
and of the average squark mass $\tilde m_Q$.}
\label{falign4}
\end{figure}

The supersymmetric contributions to $B^0-\overline{B^0}$ mixing can be 
calculated along the lines described in section III.A. The various 
contributions are proportional to $[(\delta^d_{LL})_{13}]^2$, 
$[(\delta^d_{RR})_{13}]^2$, $(\delta^d_{LL})_{13}(\delta^d_{RR})_{13}$,
$[(\delta^d_{LR})_{13}]^2$, $[(\delta^d_{RL})_{13}]^2$ and 
$(\delta^d_{LR})_{13}(\delta^d_{RL})_{13}$. For each of these contributions,
we find the value of the $(\delta^d_{MN})_{13}$ parameter that would saturate
the experimental upper bound on $|M_{12}^B|$ from eq. (\ref{delmb}), 
$|M_{12}^B|\lsim1.7\times10^{-10}\ MeV$. The results of this analysis are shown
in Fig. \ref{falign4}. For example, for $m_{\tilde g}=\tilde m_Q=1\ TeV$, we 
find that supersymmetric contributions would saturate $\Delta m_B$ if at least 
one of the following conditions is satisfied:
\beqa\label{boullot}
(\delta^d_{LL})_{13}&\sim&0.2,\no\\
\sqrt{(\delta^d_{LL})_{13}(\delta^d_{RR})_{13}}&\sim&0.04.
\eeqa

We should now compare these results to the predictions given in Table I:
\beqa\label{dotali}
(K^d_{LL})_{13}&\sim&|V_{ub}|\sim0.004,\no\\
\sqrt{(K^d_{LL})_{13}(K^d_{RR})_{13}}&\lsim&\lambda\sqrt{m_d/m_b}
\sim0.01.
\eeqa
We obtain the following approximate range for the supersymetric contribution to
$M_{12}^B$:
\beq\label{motbs}
\lambda^4\lsim\left|{(M_{12}^B)^{\rm SUSY}\over (M_{12}^B)^{\rm EXP}}
\right|\lsim\lambda^2.
\eeq
In particular, the supersymmetric contribution to the $B^0-\overline{B^0}$ 
mixing amplitude $M_{12}^B$, and hence to $\Delta m_B$ and to $a_{\psi K}$, 
is at most a few percent. 

\subsection{$B_s-\overline{B_s}$ mixing}
Within the Standard Model, the ratio between the mass differences in the
$B_s$ and $B^0$ systems, $\Delta m_{B_s}/\Delta m_B$, depends on the 
CKM elements (up to SU(3) breaking effects of order twenty percent),
\beq\label{dmbdmssm}
{\Delta m_{B_s}\over\Delta m_B}\sim\left|{V_{ts}\over V_{td}}\right|^2\sim
{1\over\lambda^2}.
\eeq
Note that $\Delta m_{B_s}$ has not been measured yet and only a lower bound
exists \cite{Groom:in},
\beq\label{dmsexp}
{\Delta m_{B_s}\over\Delta m_B}\gsim 30.
\eeq

The prediction of alignment models can be read from Table I. The relevant
ratios are
\beqa\label{mdmssusy}
{(K^d_L)^2_{23}\over(K^d_L)^2_{13}}\sim{1\over\lambda^2};\no\\ 
{\max\left[(K^d_L)_{23}(K^d_R)_{23}\right]\over
\max\left[(K^d_L)_{13}(K^d_R)_{13}\right]}\sim{1\over\lambda^2}.
\eeqa
Based on these results, we conclude that
the supersymmetric contribution to $B_s-\overline{B_s}$ mixing is at most
of order a few percent. Such an effect is too small to be clearly observed 
through a measurement of $\Delta m_{B_s}$. However, the Standard
Model prediction for the CP asymmetries in $B_s$ decay to $\psi\phi$ (or in
any other $b\to c\bar cs$ process leading to a final CP eigenstate) is of order
$\lambda^2$; these predictions can then be violated in a significant way.

\subsection{$b\to X\gamma$}
Within our framework, the structure of $\tilde M^{2q}_{LR}$ is similar to that
of $M^q$: the same holomorphic zeros appear in both, and the same parametric
suppression holds for the non-vanishing entries (though the coefficients of
order one are, in general, different). Consequently, alignment models predict 
also the parametric suppression of the chirality-changing couplings, 
$(\delta^d_{MN})_{ij}\equiv(V^d_M \tilde M^{2d}_{MN} V^{d\dagger}_N)_{ij}/
\tilde m^2$ with $M\neq N$. These predictions are given in Table II.

\begin{table}[tbh]
\begin{tabular}{c|c}
$(\delta^d_{MN})_{ij}$ &  Prediction                 \\ \hline\hline
$(\delta^d_{LR})_{12}$ & $\lambda^7\ (m_b/\tilde m)$ \\ 
$(\delta^d_{RL})_{12}$ & $\lambda^9\ (m_b/\tilde m)$ \\ \hline
$(\delta^d_{LR})_{13}$ & $\lambda^3\ (m_b/\tilde m)$ \\ 
$(\delta^d_{RL})_{13}$ & $\lambda^7\ (m_b/\tilde m)$ \\ \hline
$(\delta^d_{LR})_{23}$ & $\lambda^2\ (m_b/\tilde m)$ \\ 
$(\delta^d_{RL})_{23}$ & $\lambda^4\ (m_b/\tilde m)$ \\ 
\end{tabular}\vspace*{4pt}
\caption{Predictions for supersymmetric chirality-changing, flavor-changing
mass insertions in models of alignment. The estimates in powers of 
$\lambda\sim0.2$ refer to our evaluation of the quark mass ratios in powers of 
$\lambda$.}
\end{table}

These predictions imply that the supersymmetric contributions to $b\to 
X_s\gamma$ are small in our framework. For example, with $\tilde m\sim500\ 
GeV$, the prediction is $(\delta^d_{LR})_{23}\sim\lambda^5$ while the 
requirement, for the supersymmetric contribution to be significant, is 
$(\delta^d_{LR})_{23}\sim\lambda^2$. Thus the modificiation of
the Standard Model prediction is of order $10^{-4}$ . Even for $\tilde m$ close
to $m_Z$, the supersymmetric contribution is below the percent level.
Similar conclusions hold for the $b\to X_d\gamma$ decay.

\section{$K$ Physics}
$K$ physics have played an enormous role in shaping our thinking on
supersymmetry breaking. The very idea of alignment comes from the strong
constraints on the soft supersymmetry breaking terms that follow from the
smallness of $K^0-\overline{K^0}$ mixing. Future developments in $K$ physics 
$-$ particularly $\varepsilon^\prime/\varepsilon$ and $K\to\pi\nu\bar\nu$ 
decays $-$ are likely to test in various ways the solutions that have been 
proposed to the supersymmetric flavor problem. As concerns $\varepsilon^\prime/
\varepsilon$, one may hope that future {\it theoretical} developments will 
allow us to tell whether indeed the standard model accounts for the measured 
value. As concerns the rare $K\to\pi\nu\bar\nu$ decays, in the future the
measurement of the charged ($K^\pm$) mode might be improved and the neutral 
($K_L$) mode might be measured, providing
important information on supersymmetric flavor and CP violation. Whether 
deviations from the standard model are found or not, the results will help in 
testing alignment.

\subsection{$\varepsilon^\prime/\varepsilon$}
Direct CP violation in $K\to\pi\pi$ decays has now been measured with
high accuracy (for a review, see \cite{Nir:2001ge} and references therein):
\beq\label{epeexp}
{\varepsilon^\prime\over\varepsilon}=(1.72\pm0.18)\times10^{-3}.
\eeq
For
\beq\label{satepe}
{\cal I}m[(\delta^d_{LR})_{12}]\sim\lambda^7
\left({\tilde m\over500\ GeV}\right),
\eeq
(and/or for a similar magnitude of ${\cal I}m[(\delta^d_{RL})_{12}]$),
the supersymmetric contribution could saturate 
$\varepsilon^\prime/\varepsilon$ \cite{Masiero:1999ub}. From Table II
we learn that the predicted size is
\beq\label{preepe}
(\delta^d_{LR})_{12}\sim\lambda^7\ \left({m_b\over\tilde m}\right).
\eeq
We learn that models of alignment cannot explain a large deviation from the 
Standard Model prediction \cite{Eyal:1999gk,Dine:2001ne}. (See, however, 
ref. \cite{Baek:1999jq} for a related model where the supersymmetric 
contribution is significant.) As mentioned above, experiments have 
determined $\varepsilon^\prime/\varepsilon$ rather accurately; the question of 
whether there is room (or even necessity) for a large supersymmetric 
contribution can only be answered if the theoretical
determination of the relevant hadronic matrix elements improves in a
significant way.

\subsection{$K\to\pi\nu\bar\nu$}
The measurement of BR($K^+\to\pi^+\nu\bar\nu$) has been recently improved
\cite{Adler:2001xv}:
\beq\label{pknnexp}
{\rm BR}(K^+\to\pi^+\nu\bar\nu)=(1.57^{+1.75}_{-0.82})\times10^{-10}.
\eeq
The supersymmetric contribution can saturate this rate if 
\cite{Nir:1997tf,Buras:1997ij,Colangelo:1998pm,D'Ambrosio:2001zh}
\beq\label{suskpnn}
(\delta^d_{LL})_{12}\sim\lambda^2
\eeq
(or if $(\delta^d_{LL})_{13}(\delta^d_{LL})_{23}\sim\lambda^2$).
Examining Table I, we learn that the relevant flavor changing couplings
are much smaller. We conclude that models of alignment cannot explain a 
large deviation from the Standard Model prediction \cite{Nir:1997tf}. 
This situation might actually be helpful in probing alignment: while it may
be difficult to be convinced of new contributions at the level of a few
percent from a direct comparison between $\Delta m_B/\Delta m_{B_s}$ or 
$a_{\psi K}$ and the standard model prediction, such deviations can be
probed by a violation of the Standard Model correlations between these
observables and the $K\to\pi\nu\bar\nu$ decay rates
\cite{Buchalla:1994tr,Bergmann:2000ak}.

\section{Alignment at High Scale}
The starting point of most previously-studied models of alignment is the 
assumption that the flavor structure of the soft supersymmetry breaking terms 
is determined solely by the selection rules related to the approximate 
horizontal symmetry. When we consider, however, a high scale of supersymmetry 
breaking, renormalization group evolution (RGE) of squark masses may induce an 
approximate degeneracy at low scale. Our purpose in this chapter is to 
investigate this effect and describe the phenomenological consequences.

\subsection{RGE-Induced Degeneracy}
The RGE effects on the Yukawa matrices are small 
\cite{Aguilar-Saavedra:1996mc}, so we need to consider only the soft
supersymmetry breaking terms. For our purposes, it is sufficient to consider 
the one-loop RG equations in the limit where all Yukawa couplings are set to 
zero \cite{Choudhury:1994pn}:
\beqa\label{olrge}
\partial_t\tilde m_a&=&-{1\over4\pi}b_a\alpha_a\tilde m_a,\no\\
\partial_t(\tilde M^2_{LL})_{ij}&=&{\delta_{ij}\over4\pi}\left(
{16\over3}\alpha_3\tilde m_3^2+3\alpha_2\tilde m_2^2+
{1\over9}\alpha_1\tilde m_1^2\right)-{m_{3/2}^2\over16\pi^2}\left[\left(
A^uA^{u\dagger}\right)_{ij}+\left(A^dA^{d\dagger}\right)_{ij}\right],\no\\
\partial_t(\tilde M^{2u}_{RR})_{ij}&=&{\delta_{ij}\over4\pi}\left(
{16\over3}\alpha_3\tilde m_3^2+{16\over9}\alpha_1\tilde m_1^2\right)
-{m_{3/2}^2\over8\pi^2}\left(A^{u\dagger}A^{u}\right)_{ij},\no\\
\partial_t(\tilde M^{2d}_{RR})_{ij}&=&{\delta_{ij}\over4\pi}\left(
{16\over3}\alpha_3\tilde m_3^2+{4\over9}\alpha_1\tilde m_1^2\right)
-{m_{3/2}^2\over8\pi^2}\left(A^{d\dagger}A^{d}\right)_{ij},\no\\
\partial_t A^u_{ij}&=&{1\over4\pi}\left(
{8\over3}\alpha_3+{3\over2}\alpha_2+{13\over18}\alpha_1\right)A^u_{ij},\no\\
\partial_t A^d_{ij}&=&{1\over4\pi}\left(
{8\over3}\alpha_3+{3\over2}\alpha_2+{7\over18}\alpha_1\right)A^d_{ij},
\eeqa
where $t=2\ln(M_{\rm S}/Q)$, $M_{\rm S}$ is the scale at which supersymmetry
breaking is communicated to the MSSM, and $b_{1,2,3}=(11,1,-3)$. The $A^q$
matrices are defined through $\tilde M^{2q}_{LR}=m_{3/2} A^q\langle\phi_q
\rangle$. The important point to notice is that the squark mass-squared 
matrices, $\tilde M^{2q}_{MM}$, get large universal contributions that are 
proportional to the gauge couplings. 

Let us take, for example, $M_{\rm S}\approx M_{\rm GUT}$ and set $Q=m_Z$ 
($t\simeq67$). Then the weak scale parameters (unprimed) can be written in
terms of the high scale parameters (primed) as follows \cite{Choudhury:1994pn}:
\beqa\label{lshsal}
(\tilde M^2_{LL})_{ij}&=&(\tilde M^2_{LL})^\prime_{ij}
+7\delta_{ij}m_{1/2}^{\prime2}-m_{3/2}^{\prime2}\left[
1.8\left(A^{u\prime}A^{u\prime\dagger}\right)_{ij}
+1.7\left(A^{d\prime}A^{d\prime\dagger}\right)_{ij}\right],\no\\
(\tilde M^{2u}_{RR})_{ij}&=&(\tilde M^{2u}_{RR})^\prime_{ij}
+7\delta_{ij}m_{1/2}^{\prime2}-3.6m_{3/2}^{\prime2}
\left(A^{u\prime\dagger}A^{u\prime}\right)_{ij},\no\\
(\tilde M^{2d}_{RR})_{ij}&=&(\tilde M^{2d}_{RR})^\prime_{ij}
+7\delta_{ij}m_{1/2}^{\prime2}-3.4m_{3/2}^{\prime2}
\left(A^{d\prime\dagger}A^{d\prime}\right)_{ij},\no\\
A^u_{ij}&=&3.7A^{u\prime}_{ij},\no\\
A^d_{ij}&=&3.6A^{d\prime}_{ij}.
\eeqa
Here $m_{1/2}^\prime$ is the average gaugino mass at the GUT scale.
Thus, RGE induces a universal contribution of order ${7\over9}m_{\tilde g}^2$
to the weak-scale squark mass-squared matrices. We used here the fact that the
RGE of gaugino masses yields $m_{\tilde g}\approx3m_{1/2}^\prime$.

We now make the crucial assumption that the structure of the
soft supersymmetry breaking terms at $M_{\rm S}$  is solely determined by
the horizontal symmetry. This assumption means that, at the high scale,
the following order of magnitude relations hold:
\beq\label{nohihi}
\tilde m_{3/2}^{\prime^2}\sim\tilde m_{1/2}^{\prime^2}
\sim(\tilde M^{2q}_{MM})^\prime_{ii};\ \ \  
A^{u\prime}_{33}\sim1;\ \ \  A^{d\prime}_{33}\sim m_b/\langle\phi_d\rangle.
\eeq
But then, at low energy, squark masses acquire approximate degeneracy. 
The estimates of the supersymmetric mixing angles in Table I correspond
in this case to the suppression of the high-scale $\delta_{ij}$ parameters. But
the weak-scale $\delta_{ij}$ parameters are now suppressed not only by the 
small mixing angles but also by squark degeneracy. 
Explicitly, the low energy $\delta_{ij}$ parameters have the following 
RGE-induced suppression factors with respect to their high energy values:
\beqa\label{ledelta}
(\delta^d_{LL})_{ij}&\approx&0.25(\delta^d_{LL})_{ij}^\prime\ \ \ 
(ij)=(12),(13),(23);\no\\
(\delta^d_{RR})_{ij}&\approx&0.15(\delta^d_{RR})_{ij}^\prime\ \ \ 
(ij)=(12),(13),(23);\no\\
(\delta^u_{MM})_{12}&\approx&0.15(\delta^u_{MM})_{12}^\prime,\ \ \ M=L,R.
\eeqa
In other words, if eqs. (\ref{nodeg}) and (\ref{nohihi}) hold at the GUT scale,
we have at the
weak scale $\Delta\tilde m^2/\tilde m^2\approx1/4\ (1/7)$ in the $\tilde d_L$
sector ($\tilde d_R$ sector and first two up squark generations).
We would like to emphasize the following three points:

(i) The milder suppression of $(\delta^d_{LL})_{ij}$ depends on our
assumption that the scale that characterizes the $A$-terms is $m^\prime_{3/2}$.
If it is smaller, the degeneracy becomes as strong as in the other sectors.
The degeneracy would be similarly enhanced if the $A$-matrices were 
{\it exactly} proportional to the corresponding Yukawa-matrices.  

(ii) The results in eq. (\ref{ledelta}) have been derived with $\tan\beta=
{\cal O}(1)$. In case that $\tan\beta\gg1$, the suppression of 
$(\delta^d_{RR})_{23}$ becomes milder: $(\delta^d_{RR})_{23}\approx0.5
(\delta^d_{RR})_{23}^\prime$  (for $\tan\beta\sim m_t/m_b$).

(iii) We used here, as an example, $M_{\rm S}\approx M_{\rm GUT}$. Lower values
of $M_{\rm S}$ correspond to weaker RGE effects and, therefore, to a milder
suppression of the flavor changing effects. For $M_{\rm S}\lsim10^9\ GeV$,
there is effectively no degeneracy and the phenomenology is the same as
in the discussion in previous sections, where alignment is the only source
of suppression of flavor changing couplings.

\subsection{Phenomenological Consequences}
The RGE-induced suppression of the flavor changing $\delta_{ij}$ parameters
in the high scale models has important phenomenological consequences. Before
we list the phenomenological implications of this class of models, let us 
point out that the predictions are here somewhat sharper. This is due to the
fact that, given our assumption (\ref{nohihi}), we can estimate 
\beq\label{hiscx}
x\equiv{m_{\tilde g}^2\over\tilde m^2}\sim{9\over7}.
\eeq
Again, we used here as our example $M_S\approx M_{\rm GUT}$.
This leaves essentially a single free parameter, say, $\tilde m$, in
any given model in this class.

(i) $D^0-\overline{D^0}$ mixing: eq. (\ref{aldel}) is now replaced 
(for $M_S\approx M_{\rm GUT}$) with 
\beq\label{alhdel}
(\delta^u_{LL})_{12}\sim0.03,
\eeq 
to be compared with the constraints of Fig. \ref{falign1} (along the curve
$m_{\tilde g}/\tilde m\sim1.1$). We can make the following statements:

(a) There is no region of parameter space that is disfavored by the
experimental upper bound on $|M_{12}^D|$. In particular, the scale of
squark and gluino masses could be as low as 300 GeV. This is true for a
supersymmetry breaking scale as low as $M_{\rm S}\sim10^{14}\ GeV$: for
$M_S\gsim10^{14}\ GeV$ our framework predicts $(\delta^u_{LL})_{12}\lsim0.05$ 
which, as can be seen in Fig. \ref{falign1}, is the upper bound for $\tilde m_q 
\sim300\ GeV$.

(b) For $\tilde m\sim1$ TeV, the supersymmetric contributions to $|M_{12}^D|$ 
are a factor of ${\cal O}(50)$ below the experimental bound. Given the expected
experimental sensitivity of future experiments, it will be impossible to 
exclude models of high-scale alignment based on non-observation of
$D^0-\overline{D^0}$ mixing.

(c) For $\tilde m\sim300$ GeV, the supersymmetric contributions to $|M_{12}^D|$ 
are a factor of ${\cal O}(3)$ below the experimental bound. It is then possible
that $D^0-\overline{D^0}$ mixing will be observed in the future.

(ii) $B^0-\overline{B^0}$ mixing: eq. (\ref{dotali}) is now replaced with 
\beqa\label{dotalih}
(\delta^d_{LL})_{13}&\sim&0.001,\no\\
\sqrt{(\delta^d_{LL})_{13}(\delta^d_{RR})_{13}}&\lsim&0.003,
\eeqa
to be compared with the constraints of Fig. \ref{falign4}. We can make the 
following statements:

(a) The supersymmetric contribution to $B^0-\overline{B^0}$ mixing is smaller
by a factor of at least 10 compared to the low-scale models of similar squark 
and gluino masses. In particular, for $\tilde m\sim1$ TeV, the modification to 
the standard model prediction for $a_{\psi K}$ is below the percent level.

(b) The fact that, in this class of alignment models, light [that is, ${\cal O}
(300\ GeV)$] squark masses are allowed, means that the maximal supersymmetric
contributions could be comparable to the maximal low-scale model predictions.
Indeed, with $\tilde m\sim300$ GeV and large $\tan\beta$ (to give minimal
suppression of $(\delta^d_{RR})_{13}$), the supersymmetric contribution could
be of ${\cal O}(0.1)$ of $M_{12}^B$. This could lead to observable modifications
of $a_{\psi K}$.

\begin{figure}[t]
\centerline{\epsfxsize=0.5\textwidth \epsfbox{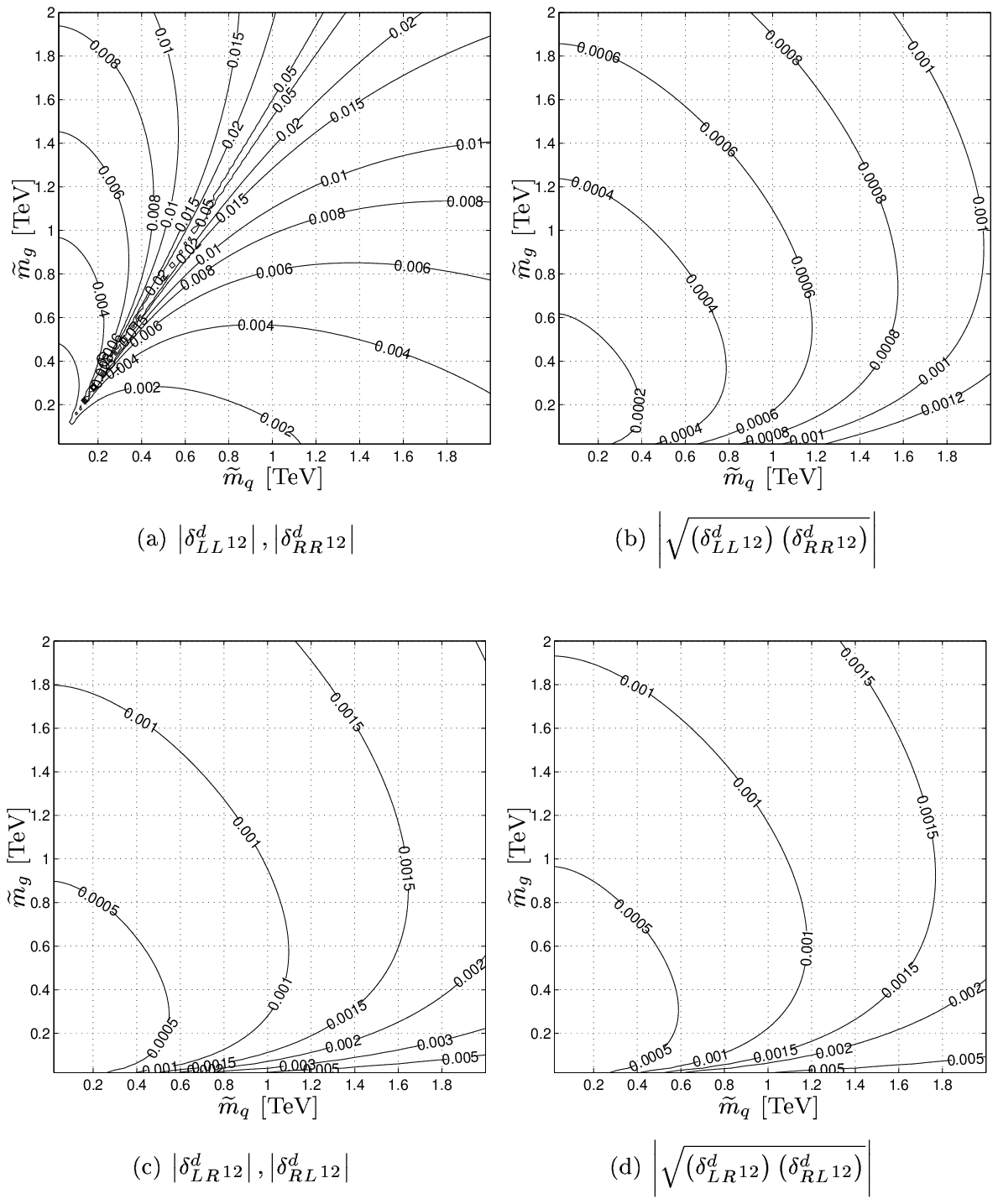}}
\caption{Constraints on flavor changing mass insertions from 
$K^0-\overline{K^0}$ mixing as a function of the gluino mass $m_{\tilde g}$ 
and of the average squark mass $\tilde m_Q$.}
\label{falign5}
\end{figure}

(iii) $K^0-\overline{K^0}$ mixing: the constraints from $\varepsilon_K$
(assuming CP violating phases of order one in the supersymmetric mixing 
matrices) are given in Fig. \ref{falign5}. For example, with 
$m_{\tilde g}=\tilde m=1$
TeV, we obtain:
\beqa\label{kkconh}
(\delta^d_{LL})_{12}&\lsim&8\times10^{-3},\no\\
\sqrt{(\delta^d_{LL})_{12}(\delta^d_{RR})_{12}}&\lsim&6\times10^{-4}.
\eeqa
Given eq. (\ref{ledelta}), these constraints can be translated into
bounds on the supersymmetric mixing angles,
\beqa\label{kkcon}
(K^d_{L})_{12}&\lsim&\lambda^2,\no\\
\sqrt{(K^d_{L})_{12}(K^d_{R})_{12}}&\lsim&\lambda^3.
\eeqa
This is to be compared with the bounds $(K^d_{L})_{12}\lsim\lambda^3$
and $\sqrt{(K^d_{L})_{12}(K^d_{R})_{12}}\lsim\lambda^5$ that apply in
low scale models of alignment.

Models of alignment are constructed to satisfy the $\Delta m_K$ and 
$\varepsilon_K$ constraints. What we have just learnt is that in models of
GUT-scale alignment, the constraints on the mixing angles (\ref{kkcon}) are
milder. The question then arises whether this situation has significant
consequences for model building.

The most dramatic result would be if the `naive' alignment,
\beq\label{naivalig}
|(K^d_L)_{12}|\sim|V_{us}|\sim\lambda,\ \ \ 
|(K^d_R)_{12}|\sim{m_d\over m_s|V_{us}|}\sim\lambda,
\eeq
were sufficient to satisfy the $K^0-\overline{K^0}$ constraints. If this were 
the case, then no holomorphic zeros would be required and the analysis of both 
model building and the phenomenological consequences of alignment would change 
considerably. What we learn from eq. (\ref{kkconh}) is, however, that this is
not the case. One could imagine that the parametric suppression gives
$|(K^d_L)_{12}|\sim\lambda$ and that an accidental suppression of ${\cal O}(6)$
would make $(\delta^d_{LL})_{12}$ consistent with the bound (\ref{kkconh}).
But then the second constraint would imply $(\delta^d_{RR})_{12}\lsim5\times
10^{-5}$, a factor of ${\cal O}(10^3)$ below the naive suppression. We conclude
that the RGE-induced suppression in the GUT-scale models is not enough to
allow viable models that employ no holomorphic zeros.

Under these circumstances, the milder constraints in eq. (\ref{kkcon}) do not
give a significant simplification for model building. In particular, relaxing
the bound on $(K^d_L)_{12}$ from $\lambda^3$ in low-scale models to $\lambda^2$
in high scale models makes no difference at all. The point is that holomorphic
zeros suppress $(K^d_L)_{12}$ compared to its naive value (\ref{naivalig})
by at least $\epsilon_{\rm max}^2$. Assuming, as we do in this work, that
$\epsilon_{\rm max}\lsim\lambda$, the consequences for model building of
the $\lambda^3$ and $\lambda^2$ bounds are identical. On the other hand,
the milder bound on $\sqrt{(K^d_L)_{12}(K^d_R)_{12}}$, $\lambda^3$ instead
of $\lambda^5$, does allow horizontal charge assignments that would not be
viable in low scale models.

We conclude that models of GUT scale alignment have phenomenological 
consequences that may be very different from low scale alignment. The
difference in model building (in the framework of Abelian horizontal
symmetries) is, however, of limited significance.

\section{Conclusions}
We analyzed questions of model building and of phenomenological
implications in the framework of quark-squark alignment.
In models of alignment, three ingredients play a role in suppressing
the supersymmetric contributions to flavor changing neutral currents:

(i) Approximate horizontal symmetries naturally suppress off-diagonal
entries in both quark and squark mass matrices. This alignment of mass
matrices induces small mixing angles in gaugino couplings.

(ii) Supersymmetry requires that the Yukawa couplings are holomorphic.
In combination with the horizontal symmetries, zero textures may be
required by holomorphy, opening up the possibility of a very precise alignment.

(iii) The running of the soft supersymmetry breaking terms may induce
approximate degeneracy among squarks, even if there is no degeneracy in
the high energy theory. 

On the model-building side, we have made the following two main points:

(a) Under a few reasonable assumptions, there is a unique phenomenologically 
viable structure for the down quark mass matrix. In particular, four 
holomorphic zeros must appear, $M_{12}^d=M_{21}^d=M_{31}^d=M_{32}^d=0$.

(b) The possibility that a certain degree of degeneracy is induced by 
RGE somewhat relaxes the constraints on the required alignment.
Still, `naive' alignment where, for example, the supersymmetric mixing
angles for doublet quarks and squarks have the same parametric suppression
as the corresponding CKM angles, is not viable. Consequently, the same 
holomorphic zeros must play a role and the complications of model building are 
not simplified.

On the phenomenological side, we would like to make the following points
regarding the future prospects for discovering or excluding the idea
of quark-squark alignment:

(a)  Alignment models without squark degeneracy require that $|M_{12}^D|$ 
should be close to present experimental bounds. If the bounds on 
$D^0-\overline{D^0}$ mixing are improved by an order of magnitude, such models 
will be disfavored. Note that to improve the bound on $M_{12}^D$
by an order of magnitude, it is not necessarily required to improve the bound
on $\Delta m_D$ by a similar factor. A mild experimental progress in 
constraining each of $x\equiv\Delta m_D/\Gamma$, $\phi_D$ (the relevant weak
phase) and $\delta$ (the relevant strong phase) might give a
significantly improved bound on $|M_{12}^D|$

(b) The supersymmetric contribution to the $B^0-\overline{B^0}$ mixing 
amplitude $M_{12}^B$ is at most a few percent of the experimental value. 
Experimentally, both $\Delta m_B$ and $a_{\psi K_S}$ can be measured with
an accuracy better than a few percent. The question of whether a deviation
of order of a few percent from the Standard Model predictions can be 
convincingly signalled is related to the theoretical accuracy of the
predictions. Given the hadronic uncertainties in the calculation of 
$\Delta m_B$, it will be impossible to have a convincing signal for this new 
contribution from the measurement of the mass difference. On the other hand, 
the hadronic uncertainties in the Standard Model relation 
$a_{\psi K}=\sin2\beta$ are smaller than a percent. It is still an open
question whether the value of $2\beta$, constrained by other measurements,
can be determined with the required accuracy.

(c) The supersymmetric contribution to the $B_s-\overline{B_s}$ mixing 
amplitude $M_{12}^{B_s}$ is at most a few percent of the experimental lower 
bound. Again, it would be difficult to have a convincing signal for this new 
contribution from the measurement of the mass difference $\Delta m_{B_s}$. On 
the other hand, the Standard Model predicts small [${\cal O}(\lambda^2)$]
CP asymmetries in $B_s$ decays to final CP eigenstates that involve the
$b\to c\bar cs$ quark subprocess, so that the deviation can be significant.

(d) The supersymmetric contributions to $K\to\pi\nu\bar\nu$ decays are
small. Thus the correlations between these decay rates and various observables
related to $B^0-\overline{B^0}$ mixing, that are cleanly predicted by the
Standard Model, may be violated.

We conclude that the observation of CP violation in $D^0-\overline{D^0}$
mixing and shifts of ${\cal O}(\lambda^2)$ from the Standard Model
predictions for CP asymmetries in $B^0$ and $B_s$ decays are the best
possible clues for alignment. On the other hand, given the possibility of 
RGE-induced approximate degeneracy, it will be difficult to exclude the idea 
of alignment if no deviations from the Standard Model are observed. Stronger 
constraints on such deviations will simply translate into stronger lower 
bounds on the scale where alignment holds.

\begin{acknowledgments}
We thank Yael Shadmi for useful discussions.
Y.N.\ is supported by the Israel Science Foundation founded by the
Israel Academy of Sciences and Humanities.
\end{acknowledgments}


\end{document}